\documentclass[12pt,letterpaper,english]{article}
\usepackage{amssymb,amsmath}
\include{epsf}
\usepackage{slashed}

\usepackage{graphicx}
\usepackage{amsfonts}

\newcommand{\ii}{{\rm i}}

\newcommand{\eqn}[1]{(\ref{#1})}

\def\appendix#1{\addtocounter{section}{1}\setcounter{equation}{0}
\renewcommand{\thesection}{\Alph{section}}
\section*{
\thesection\protect\indent \parbox[t]{11.715cm} {#1}}
\addcontentsline{toc}{section}{Appendix\thesection\ \ \ #1} }
\newcommand{\complex}{{\mathbb C}} 

\newcommand{\tr}[1]{\:{\rm tr}\,#1}
\newcommand{\Tr}[1]{\:{\rm Tr}\,#1}
\newcommand{\be}{\begin{equation}}
\newcommand{\ee}{\end{equation}}
\newcommand{\beq}{\begin{equation}}
\newcommand{\eeq}{\end{equation}}
\newcommand{\bea}{\begin{eqnarray}}
\newcommand{\eea}{\end{eqnarray}}
\def\beqa{\begin{eqnarray}}
\def\eeqa{\end{eqnarray}}
\def\nn{\nonumber}

\newcommand{\eq}{\begin{equation}}
\newcommand{\eqa}{\begin{eqnarray}}
\newcommand{\en}{\end{equation}}
\newcommand{\ena}{\end{eqnarray}}

\newcommand{\X}{\mathfrak X}

\def\R{{\mathbb R}}  
 \def\one{\mbox{1 \kern-.59em {\rm l}}}

\def\a{\alpha} 
 \def\L{\Lambda} \def\la{\lambda}

\def\cA{{\cal A}}      
\def\cH{{\cal H}}      
  \def\cQ{{\cal Q}}

\def\obar{\overline}

\begin{document}
\begin{titlepage}
\begin{flushright}

\baselineskip=12pt
DSF/x/2010\\
ICCUB-10-xxx\\
UWTHPh-2010-x\\ \hfill{ }
\end{flushright}

\begin{center}

{\Large\bf Gauge Symmetry Breaking in Matrix Models}\footnote{Talk delivered by
F.~Lizzi at the 2nd School on Quantum Gravity and Quantum Geometry session of
the 9th Hellenic School on Elementary Particle Physics and Gravity, Corfu
2009.}

\baselineskip=14pt

\vspace{1cm}

{\bf Harald Grosse$^{a}$,  {\bf Fedele Lizzi\,}$^{b,c}$ and {\bf
Harold Steinacker}$^{a}$}
\\[5mm] $^a$ {\it Department of Physics, University of Vienna\\Boltzmanngasse5, A-190 Vienna, Austria}
\\{\small \tt
harald.grosse, harold.steinacker @univie.ac.at}\\[6mm]
$^b$ {\it Dipartimento di Scienze Fisiche, Universit\`{a} di
Napoli {\sl Federico II}} and {\it INFN, Sezione di Napoli,
Via Cintia, 80126 Napoli, Italy}\\
{\small \tt fedele.lizzi@na.infn.it}
\\[5mm]
$^c$ {\it High Energy Physics Group, Dept.\ Estructura i
Constituents de la Mat\`eria and Institut de Ci\`encies del Cosmos
\\Universitat de Barcelona Barcelona, Catalonia, Spain}

\end{center}


\begin{abstract}
We argue that some features of the standard model, in particular the
fermion assignment and symmetry breaking, can be obtained in matrix
model which describes noncommutative gauge theory as well as gravity
in an emergent way. The mechanism is based on the presence of some
extra (matrix) dimensions. These extra dimensions are different from
the usual ones which give to a noncommutative geometry of the
Gr\"onewold-Moyal type, and are reminiscent of the Connes-Lott
model, although the action is very different.

\end{abstract}

\end{titlepage}

\section{Introduction}
Matrix models such as the ones introduced
in~\cite{EguchiKawai,IKKT,Alekseev:2000fd} and noncommutative
geometry~\cite{connes,landi,ticos,Madore} may be the appropriate
tools to describe physics in the quantum gravity regime, where the
ordinary concepts of manifold may no longer be valid. The matrix
models are known to describe noncommutative gauge
theory~\cite{Aoki:1999vr,Douglas:2001ba}, and contain gravity as
an emergent phenomenon~\cite{Steinackeroriginal} a la
Sakharov~\cite{Sakharov, Visser}. On the other side noncommutative
geometry can describe the Higgs mechanism of symmetry
breaking~\cite{ConnesLott, Dubois-VioletteKernerMadore,
ScheckCoquerauxEsposito-Farese} and more in general the standard
model coupled with gravity~\cite{connes, ConnesLott, AC2M2} in a
natural way by extending the space with the addition of some extra
(in general noncommuting) coordinates, while the rest of the
coordinates remain the same.

We will describe a construction presented
in~\cite{GrosseLizziSteinacker} (see also~\cite{Wroclawproc})
which is able to reproduce some of the features of the standard
model, mainly in relation to symmetry breaking. 
The symmetry breaking happens through VEV's in the extra
coordinates analogous to the ideas in \cite{Aschieri:2006uw}, 
and all fermions of the standard model are accommodated.
The model
presented here is not yet fully realistic, and at present it has
no phenomenological valence, it nevertheless shows that matrix
models in the context of noncommutative geometry has the
potentiality to describe models which in the future may have an
actual predictive power.

\section{The matrix Model}

\setcounter{equation}{0}

The starting point is the action:
\be
S_{YM} = - (2\pi)^{2}\frac{\Lambda_{NC}^4}{g^2}\, \Tr
\left([X^a,X^b] [X^{a'},X^{b'}] \eta_{aa'}\eta_{bb'} \,\, + \,\,
\obar\Psi  \Gamma_a [X^a,\Psi] \right) \label{YM-action-1}
\ee
where $X^a$ are infinite-dimensional hermitian matrices,
$\eta_{aa'}$ is the flat Minkowski (or Euclidean) metric and
$\Psi$ is a corresponding (Grassmann-valued) spinor which is also
an infinite matrix, $g$ is a couling constant and $\Lambda$ is a
an energy sclae which plays the role of noncommutativity scale. We
will not specify further which type of matrices we are dealing
with, since we are describing a crude approximation of a more
refined (and yet unknown) mathematical structure.

Consider first the bosonic part of the model, the equations of
motion for the $X$'s are
\be
[X^a,[X^b,X^{a'}]]\eta_{aa'} = 0 \label{eqmotion}
\ee
and the model is invariant under the symmetry
\be
X^a \to U X^a U^{-1} , \qquad U \in U(\cH) .
\ee
Apart form the null solution or the case in which all matrices
commutes, there is an important solution to the equations of
motion:
\be
[X_0^a,X_0^b]=i\theta^{ab} \label{u1vacuum}
\ee
with $\theta^{ab}$ constant. We call this solution the ``scalar
Moyal-Weyl'' vacuum since in this case the $X$'s are the
generators of an algebra which is isomorphic (under appropriate
regularity conditions) to the algebra of functions multiplied with
the Gr\"onelwol-Moyal $\star$-product. This matrix model describes
a noncommutative space with a constant commutator. To a first
approximation this is true for sufficiently short distance scale.
We will later consider the matrix model obtained by letting this
background fluctuate. There is a gauge invariance which at first
sight appears to be the unitary group of matrices functions of
$X$, but as shown in~\cite{Steinackeroriginal} the $U(1)$ part of
the group contributes to the gravitational degree of freedom, we
refer to the original paper for details.

The Moyal vacuum is of course not the only minimum of the action,
for example
\be
\bar X^a=X_0^a\otimes\one_n \label{XNtimes}
\ee
is another solution of~\eqn{eqmotion}, which correspond  a
noncommutative $U(n)$ gauge symmetry because in the semiclassical
limit it corresponds to a nonabelian gauge theory. Again the
$U(1)$ degree of freedom is absorbed by gravity and therefore the
theory corresponds to a $SU(n)$ gauge theory.

We can also consider the possibility that some of the dimension are
of a different kind altogether, so to have a four dimensional
spacetime which in some commutative limit goes to the usual
Minkowskian space, plus an internal space described by the tensor
product of finite dimensional matrices times the identity. This is
actually a noncommutative version of the programme that Connes and
collaborators have been developing for some time to describe the
standard model. The origins of the model lie in the aim to describe
the Higgs mechanism by some noncommutative internal coordinates,
which in some cases are taken to be
fermionic~\cite{Dubois-VioletteKernerMadore,
ScheckCoquerauxEsposito-Farese}. In the original Connes-Lott
model~\cite{ConnesLott} model the internal coordinates were two by
two diagonal matrices, but then in the evolutions of the
model~\cite{Connesreal, spectralaction, AC2M2} the internal space is
described by the matrix algebra $M_3(\complex)\times\mathbb H\times
\complex$, with $M_3(\complex)$ the algebra of three by three
complex valued matrices, and $\mathbb H$ the algebra of quaternions.
The unimodular part of this algebra corresponds to the standard
model group. The action is composed of two parts, the fermionic
action is the usual one, while the bosonic action is a regularized
version of the trac of the covariant Dirac operator. It is actually
possible to derive the bosonic action from the fermionic one
imposing scale invariance and demanding cancellation of the
anomalies~\cite{AndrianovBonora, AndrianovLizzi}. In our case the
action is different, but the idea is similar, i.e.\ we consider the
case for which the dimensions come in two kinds,
\be
X^a = (X^\mu,\X^i), \qquad \mu = 0,...,3,\,\, i = 1,..., n
\ee
with the  $X^\mu$'s  (quantized) coordinate functions of the
form~\eqn{XNtimes}, which generate the Moyal-Weyl plane, and $n$
extra generators $\X^i$. These coordinate are a solution
of~\eqn{eqmotion} because
\be
\, [\bar X^\mu,\bar X^\nu] = \ii\theta^{\mu\nu}\otimes\one_N. \qquad
[\bar X^\mu,\bar \X^i] = 0 \label{barXmu}
\ee
with $\theta$ constant. The symmetry is still $SU(N)$.

The fluctuations around this solution can be expressed in terms of
two fields, $\cA$ and $\Phi$ and the noncommutativity scale
$\Lambda_{NC}$ as
\be
X^\mu = \bar X^\mu + \cA^\mu,
\qquad     \Phi^i =   \Lambda_{NC}^2\,\X^i \label{barXmufluct}
\ee
As mentioned the trace-$U(1)$ give rise to emergent gravity and
will be ignored in the rest of these proceedings. The remaining
$SU(N)$-valued fluctuations
\be
\cA^\mu = - \theta^{\mu\nu}A_\mu^\a(x) \otimes\la_\a
\ee
correspond to $SU(N)$-valued gauge fields, while the fluctuations
in the internal degrees of freedom
\be
\Phi^{i} = \Phi^{i,\a}(x) \otimes\la_\a
\ee
correspond to scalar fields in the adjoint. The matrix model
action~\eqn{YM-action-1} therefore describes $SU(N)$ gauge theory
on $\R^4_\theta$ coupled to $n$ scalar fields. Hereafter we drop
the $\otimes$ sign.

Noncommutative gauge theory is obtained from the matrix model
using
\bea
\,[\bar X^\mu + \cA^\mu,f] &=& i \theta^{\mu\nu}
(\frac{\partial}{\partial\bar x^\nu}
+ i [A_\nu,.]) f \, \equiv \, i \theta^{\mu\nu} D_\nu f .
\eea
The matrix model action~\eqn{YM-action-1} can then be written as
\bea
S_{YM} &=& \frac 1{g^2}\,\int d^4 \bar x\,
\tr\Big( G^{\mu\mu'}\,G^{\nu\nu'}\,F_{\mu\nu}\,F_{\mu'\nu'}  \nn\\
&& \quad + 2\,G^{\mu\nu}\, D_\mu\Phi^i D_\nu \Phi^i \delta_{ij}
- [\Phi^i,\Phi^j][\Phi^{i'},\Phi^{j'}] \delta_{ij} \delta_{i'j'} \nn\\
&& \quad + \bar\Psi \slashed{D} \Psi
 + \obar\Psi \Gamma_i[\Phi^i,\Psi] \Big) .
\label{YM-action-2}
\eea
This is the action of a $SU(N)$ gauge theory on $\R^4_\theta$,
with effective metric given by
\be
G^{\mu\nu} = \rho\theta^{\mu\mu'}\theta^{\nu\nu'}\eta_{\mu'\nu'},
\qquad \rho = (\det\theta^{\mu\nu})^{-1/2} = \Lambda_{NC}^4,
\label{G-def}
\ee
which satisfies $\sqrt{|G|}=1$. Here $F_{\mu\nu} = \partial_\mu
A_\nu - \partial_\nu A_\mu + i[A_\mu,A_\nu]\,$ is the field
strength on $\R^4_\theta$ and $D_\mu \equiv \partial_\mu + i
[A_\mu,.]$ is the covariant derivative for fields in the adjoint,
and $\tr()$ denotes the trace over the $SU(N)$ components.
 The effective Dirac operator
is given by
\be
\slashed{D}\Psi = \Gamma_\mu \left[X^\mu, \Psi\right]
\,\sim\,  i\gamma^\mu D_\mu \Psi
\label{eq:Dirac op}
\ee
where~\cite{KlammerSteinackerfermions}
\be
\gamma^\mu = \sqrt{\rho}\,\Gamma_\nu \theta^{\nu\mu},
\qquad \{\gamma^\mu,\gamma^\nu\} = 2 G^{\mu\nu}.
\ee
The fermions have been rescaled appropriately,
and a constant shift as well as total derivatives
in the action are dropped.
Note that $g$ is now identified as the coupling constant
for the nonabelian gauge fields on $\R^4_\theta$.

\section{Symmetry breaking \label{se:symmbreakextradim}}
\setcounter{equation}{0}

In~\cite{GrosseLizziSteinacker} we presented two mechanisms for the
breaking of the symmetries, one based on constant matrices and the
other on seeing the internal space as fuzzy spheres. For reasons of
space in this proceedings we will present only the former, which can
be also seen as an effective version of the latter.

Consider a model with $X^\mu$ as in~\eqn{XNtimes} with $n=7$ and one
extra coordinate:
\be
\langle\X^\Phi\rangle =\left(\begin{array}{ccc}
\alpha_1\one_{2}& \  & \ \\ \ & \alpha_2\one_{2} & \ \\
\ & \ & \alpha_3\one_{3} \end{array}\right) .
\label{X-phi-vac}
\ee
The $\alpha$'s are constant quantities with the dimensions of a
length, all different among themselves. These new coordinates are
still solutions of the equations of the motion because, but the
$SU(7)$ symmetry is broken down because of $\X^\Phi$. The residual
unbroken group is $SU(3)\times SU(2) \times U(1) \times U(1) \times
U(1)$.


In the bosonic action as the spacetime ($\mu\nu$) part of action
remains unchanged, while for the $\mu\phi$ components we obtain, in
the Moyal-Weyl background,
\bea
[\bar X^\mu +\cA^\mu, \X^\Phi] &=& i\theta^{\mu\nu} D_\nu \X^\phi
=  i\theta^{\mu\nu} (\partial_\nu +  i A_\nu) \X^\phi , \nn\\
-(2\pi)^2 \Tr [X^\mu,\X^\phi] [X^{\nu},\X^\phi] \eta_{\mu\nu} &=&
\int d^4 x G^{\mu\nu} \left(\partial_\mu \X^\Phi \partial_\nu
\X^\Phi
- [A_\mu,\X^\Phi][A_\nu,\X^\Phi]\right) . \nn\\
\label{kinetic-X-A}
\eea
Note that the mixed terms $\int \partial^\mu \X^\Phi [A_\mu,\X^\Phi]
= - \frac 12 \int  \X^\phi [\partial^\mu A_\mu,\X^\Phi] = 0$
vanish, assuming the Lorentz gauge $\partial^\mu A_\mu = 0$.

Now consider the vacuum~\eqn{X-phi-vac}. Since $X^\mu$ and
$\langle\X^\Phi\rangle$ commute, this means $\langle\X^\Phi\rangle =
\mathrm{const}$ and the first term in the integral above vanish. We
can therefore separate the fluctuations of this extra dimension
which are a field, the (high energy) Higgs field. In the action the
first term is nothing but the derivative of it. The second term
instead is
\be
[A^\mu,\langle\X^\Phi\rangle]=\left(\begin{array}{ccc}
 0 & (\alpha_2-\alpha_1) A^\mu_{12} & (\alpha_3-\alpha_1) A^\mu_{13} \\
 (\alpha_1-\alpha_2) A^\mu_{21} & 0 & (\alpha_3-\alpha_2) A^\mu_{23} \\
 (\alpha_1-\alpha_3) A^\mu_{31} & (\alpha_2-\alpha_3) A^\mu_{32} & 0
\end{array}\right)
\ee
where we consider the block form of $A^\mu$
\be
A^\mu=\left(\begin{array}{ccc}
 A^\mu_{11} & A^\mu_{12} &  A^\mu_{13} \\
 A^\mu_{21} & A^\mu_{22} &  A^\mu_{23} \\
 A^\mu_{31} &  A^\mu_{32} & A^\mu_{33}
\end{array}\right)
\ee
Therefore~\eqn{kinetic-X-A} leads to the mass terms for the
off-diagonal gauge fields,
\be
-(2\pi)^2 \Tr [X^\mu,\langle\X^\Phi\rangle] [X^{\nu},\langle\X^\Phi\rangle] \eta_{\mu\nu}
=\int d^4 x G^{\mu\nu}
\left(\sum (\a_i-\a_j)^2 A_{\mu,ij} A_{\nu,ji} \right)
\label{A-mass-term}
\ee
which is nothing but the usual Higgs effect. If the differences
$\alpha_i-\alpha_j$ are large (which we assume) the non diagonal
blocks of $A^\mu$ acquire large masses $m_{ij}^2 \sim (\a_i-\a_j)^2
$, and effectively disappear from the spectrum.

\section{Particles and symmetries}

\setcounter{equation}{0}

We now show how the fermions in the standard model can be naturally
accommodated in the framework of matrix models. This is nontrivial
because the  fermions in the matrix model are necessarily in the
adjoint of some basic $SU(N)$ gauge group.
In~\cite{GrosseLizziSteinacker} have also show how the electroweak
symmetry can be broken through a somewhat modified Higgs sector, and
the Yukawa couplings which are are obtained.

For the sake of this paper we accomodate all known fermions (with
the exception of right handed neutrinos) in an upper triangular
matrix\footnote{Some of the zero's in the matrix may correspond to
other particles, see~\cite{GrosseLizziSteinacker}.}:
\be
\Psi= 
\begin{pmatrix}
  0_{2\times 2} & L_L & Q_L \\
  0_{2\times 2} & {\scriptsize \begin{array}{cc}
  0 &  e_R \\
  0 & 0
\end{array}} & Q_R \\
  0_{3\times 2} &  0_{3\times 2} & 0_{3\times 3}
\end{pmatrix}
\label{particle-assign}
\ee
Here $l_L$ will be the standard (left-handed) leptons, and $e_R$ the
right-handed electron.  The quark matrix is
\bea
\cQ &=&  \begin{pmatrix}
  Q_L \\
  Q_R \end{pmatrix} , \nn\\
Q_L &=& \begin{pmatrix}
   u_L \\ d_L
\end{pmatrix}, \qquad
Q_R = \begin{pmatrix}
   d_R \\ u_R
\end{pmatrix}
\eea

The correct hypercharge, electric charge and baryon number are
then reproduced by the following traceless generators
\bea
Y &=& \begin{pmatrix}
   0_{2\times 2}  &  &  &  \\
   & - \sigma_3 & &  \\
   & &  & -\frac 13 \one_{3\times 3}
\end{pmatrix} +\frac17 \one   \label{Ygenerator}\\
Q &=& T_3 + \frac Y2 = \frac 12 \begin{pmatrix}
   \sigma_3  &  &  &  \\
   & - \sigma_3 & &  \\
   & &  & -\frac 13 \one_{3\times 3}
\end{pmatrix} +\frac1{14} \one  \label{Qgenerator}\\
B &=& \begin{pmatrix}
   0  &  &  &  \\
   & 0 & &  \\
   & &  & - \frac 13\one_{3\times 3}
\end{pmatrix} +\frac17 \one \label{Bgenerator}
\eea
which act in the adjoint. Of course at this stage we are still quite
far from a complete model. Nevertheless this result points to the
possibility to describe the standard model within this
noncommutative geometry matrix model.

\section{Electroweak breaking}

Now we show how electroweak symmetry breaking might be realized in
this framework. To explain the idea we will first present a
simplified version where the Higgs is realized in terms of a single
extra coordinate (resp.\ scalar) field. This is again not intended
as a realistic model, but it shows that suitable Higgs potential can
naturally arise within the present framework.

Higgs field connects the left with the right sectors of leptons, and
is otherwise colour blind, it is therefore natural to consider
another extra coordinate which will have to necessarily be
off-diagonal. The following matrix has the correct characteristics:
\be
\X^\phi=\L_{NC}^{-2}\left(\begin{matrix}
  0_{2\times 2} & \phi & 0_{2\times 3} \\
  \phi^\dagger & 0_{2\times 2} & 0_{2\times 2} \\
  0_{3\times 2} &  0_{3\times 2} & 0_{3\times 3}
\end{matrix}\right)
\label{Higgs-coord}
\ee
again we consider the extra variable $\X$, its vacuum expectation
value and the fluctuations which are a physical field. The Higgs
$\phi$ is a $2 \times 2$ matrix which is actually composed of two
doublets:
\be
\phi = \begin{pmatrix} \tilde \varphi, & \varphi
\end{pmatrix} \label{two-higgs}
\ee
The vacuum expectation value of $\phi$ is an off-diagonal matrix:
\be
\langle\phi\rangle=\left(\begin{matrix} 0 & v\\ \tilde v & 0\end{matrix}\right)
\ee
All other components are assumed to be very massive, e.g. due to the
commutator with the high-energy breaking discussed before.

Now consider the fermionic part of the action~\eqn{YM-action-1}, which can be
written on $\R^4_\theta$ in the form ~\eqn{YM-action-2}. The part involving
$X^\mu$ gives the usual Dirac action as in~\eqn{YM-action-2}, and the part
involving $\X^\phi$ yields the Yukawa couplings
\be
S_Y = \Tr \obar\Psi \gamma_5 [\X^\phi,\Psi]
\ee
giving mass to the fermions. here we have considered the extra dimension to be
a fifth dimension, hence the presence of $\gamma_5$, and
$\obar\Psi=\Psi^\dagger \gamma_0$. Then the full Yukawa term is
\be
S_Y = \Tr  \obar L_L \gamma_5 \phi \left({\scriptsize
\begin{array}{cc}
  0 &  e_R \\
  0 & 0
\end{array}}\right)
+ \left({\scriptsize \begin{array}{cc}
  0 &  0 \\
 \obar e_R & 0
\end{array}}\right) \gamma_5  \phi^\dagger L
 + \obar Q_L \gamma_5 \phi Q_R
+ \obar Q_R \gamma_5 \phi^\dagger Q_L \Big) \label{Yukawa-general}
\ee
Only the correct couplings appear, albeit all with the same value.

\section{Conclusions and Outlook}

We have sketched how a matrix model with the capability to describe
noncommutative geometry and gravity may also describe some features
of the standard model, mostly regarding the issues of symmetry
breaking. The model described here is in its simplest form and is
not yet phenomenologically viable, it just points the way to
further developments. Some further steps were already undertaken
in~\cite{GrosseLizziSteinacker,Aschieri:2006uw} 
were it shown that already seeing
the extra dimensions as composed of fuzzy spheres gives more liberty
in the model, while (near-)realistic models 
appear possible along the
lines of \cite{orbifolds-forth}. 
The hope is that matrix model can not only describe
some form of quantum gravity, but also give some imput as far gauge
theories are concerned.

\paragraph{Acknowledgements} {\small We thank J.~Barrett, L.~Jonke and
G.~Zoupanos for organizing a most interesting meeting. FL~would like to thank
the Department of Estructura i Constituents de la materia, and the Institut de
Ci\`encies del Cosmos, Universitat de Barcelona for hospitality. His work was
supported in part by CUR Generalitat de Catalunya under project 2009SGR502. The
work of H. S. was supported in part by the FWF projects P20017 and  P21610.}

\end{document}